\documentclass[
  reprint,
  amsmath,
  amssymb,
  aps,
]{revtex4-2}

\usepackage[utf8]{inputenc}

\usepackage{amsfonts}
\usepackage{amsthm}
\usepackage{mathtools}
\usepackage{bm}
\usepackage{braket}

\newtheorem{theorem}{Theorem}
\newtheorem{proposition}{Proposition}

\newcommand{\mL}{\mathcal{L}}
\DeclareMathOperator{\Tr}{Tr}

\usepackage{graphicx}
\usepackage{xcolor}

\newif\ifmarkup
\markupfalse
\definecolor{revcolor}{RGB}{190,70,0}
\colorlet{REVCOLOR}{revcolor}  
\newcommand{\rev}[1]{\ifmarkup{\color{revcolor}#1}\else#1\fi}
\newenvironment{revblock}{\ifmarkup\color{revcolor}\fi}{}
\newcommand{\revfloat}{\ifmarkup\color{revcolor}\fi}  
\definecolor{cutcolor}{RGB}{110,60,160}

\newcommand{\revhead}{\ifmarkup\color{revcolor}\fi}
\newtheoremstyle{revplain}{\topsep}{\topsep}{\itshape}{}{\revhead\bfseries}{.}{ }{}
\theoremstyle{revplain}
\newtheorem{revproposition}[proposition]{Proposition}
\theoremstyle{plain}

\usepackage{tikz-cd}
\usepackage[justification=raggedright,singlelinecheck=false]{caption}
\usepackage{subcaption}

\usepackage{dcolumn}   
\usepackage{booktabs}  

\usepackage{soul}      
\usepackage{lipsum}    

\usepackage[
  colorlinks=true,
  citecolor=blue,
  urlcolor=blue,
  linkcolor=blue
]{hyperref}

\begin{document}

\preprint{APS/123-QED}

\title{\texorpdfstring{Uniformly Convergent Perturbation Theory for \rev{Open Quantum Systems} \rev{at,} near\rev{,} and away from Exceptional Points}{Uniformly Convergent Perturbation Theory for \rev{Open Quantum Systems} at, near, and away from Exceptional Points}}

\author{Pedro Linck}
\email{Contact author: pedro.linck@ufpe.br}
\affiliation{Departamento de Física, Centro de Ciências Exatas e da Natureza, Universidade Federal de Pernambuco, Recife-PE, Brazil, 50670-901}
  


\date{\today}

\begin{abstract}
Near a Liouvillian exceptional point, spectral perturbation theory becomes
ill-conditioned, yet the observable dynamics of a finite-dimensional open system
remains regular in the perturbation. Using a Dyson expansion, which requires no
diagonalization of the perturbed generator, we analyze the scaling, stability,
and convergence of the perturbative corrections \rev{at,} near\rev{,} and away
from exceptional points. We derive scaling laws showing that nonperipheral
Jordan blocks \rev{raise the degree of the} polynomial-in-time factors
multiplying decaying exponentials\rev{, which are present for any generator as
resonances, and we quantify this excess}. We further establish a
uniform-in-time error estimate for the perturbative series, valid for any
relaxing finite-dimensional Lindbladian and controlled by \rev{a parameter $q = |\varepsilon|R$, where $\varepsilon$ is the perturbation strength and $R$ is a constant built from unperturbed data. All constants of the bound remain continuous through an
exceptional point when the coalescing eigenvalues are kept in one spectral
cluster. The estimate extends to the frequency
domain via the Laplace transform, and to time-dependent perturbations. We test it in a driven
two-level system with a Liouvillian exceptional point, including perturbations
that break its symmetry.}
\end{abstract}

\maketitle

\section{\label{sec:introduction}Introduction}

Open quantum systems provide the standard framework for quantum dynamics in
the presence of dissipation, noise, and environmental
monitoring~\cite{Gorini_1976,Lindblad_1976,Breuer_2002}, with applications in
quantum optics~\cite{Fitzpatrick2017DissipativePhaseTransitionCircuitQED,
HwangRablPlenio2018DissipativeRabi, CarusottoCiuti2013QuantumFluidsLight},
condensed matter physics~\cite{SiebererBuchholdMarinoDiehl2025UniversalityOpenQuantumMatter,
SkinnerRuhmanNahum2019MeasurementInducedPhaseTransitions,
Bardyn2013TopologyByDissipation}, and decoherence-based approaches to quantum
cosmology and semiclassical gravity~\cite{LittreMatacz1993QuantumNoise,
GellMannHartle1993ClassicalEquations, GellMannHartle1990QuantumMechanicsCosmology}.
Their generators are generally non-Hermitian superoperators acting on density
matrices, and this non-Hermitian structure governs relaxation, decay modes,
dissipative gaps, transient amplification, and the approach to nonequilibrium
steady states~\cite{Breuer_2002}. Perturbation theory is the natural tool for
dynamics that deviate slightly from a known one. Existing approaches include
spectral perturbation theory~\cite{Kato_1995,li2014}, perturbative
constructions for Lindblad master equations~\cite{Yi_2000,
VillegasMartinez2016PerturbationMasterEquation, Shishkov_2020}, and response
theory for open quantum systems~\cite{Wei2011LinearResponseOpenSystems,
Levy_2021}.

Open quantum systems also exhibit exceptional points, where eigenvalues and
eigenvectors coalesce and the generator becomes
defective~\cite{Heiss_2012,Miri_2019,Ashida_2020}. Coalescences of the full
Liouvillian spectrum, the Liouvillian exceptional points, are in general
distinct from those of the underlying non-Hermitian Hamiltonian once quantum
jumps are taken into account~\cite{Minganti_2019,Minganti_2020,Sun_2024}.
Exceptional points have implications for sensing and optical
platforms~\cite{Miri_2019}, driven-dissipative
systems~\cite{CamposVenuti_2016}, relaxation
phenomena~\cite{Zhou2023AcceleratingRelaxation}, and quantum heat
machines~\cite{Alicki_1979,Kosloff_2013,Vinjanampathy_2016,Goold2016,Binder_2018,Myers2022},
where they have been used for dynamical control~\cite{Zhang_2022} and to
enhance heat-engine performance by encircling a Liouvillian exceptional
point~\cite{Bu_2023}.

Near an exceptional point, spectral perturbation theory becomes
ill-conditioned: the perturbed eigenvalues acquire the fractional-power
(Puiseux) branches characteristic of defective
degeneracies~\cite{Seyranian_2005}. \rev{The dynamics itself, however,
remains regular: for a finite-dimensional generator, the propagator at any
fixed time is an entire function of the perturbation strength.} Moreover, the
relevant question is often not how each perturbed eigenvalue behaves, but how
a weak perturbation changes populations, coherences, currents, heat flows, or
work output over a finite time window, as in response-theoretic treatments of
open systems near exceptional points~\cite{Albert_2016,Hashemi_2022,Wiersig_2022}.
This calls for a perturbation theory organized at the level of density
matrices and observables rather than around defective spectral data.
\rev{We therefore use the Dyson--Duhamel series of semigroup perturbation
theory~\cite{EngelNagel2000,Brendle2001}, which builds the corrections to the
density matrix and to expectation values directly from the unperturbed
propagator.}

\begin{revblock}
Our main results are the following. Polynomial-in-time factors appear in the
expansion for any generator, through resonances between repeated eigenvalues
along a perturbative path. Proposition~\ref{prop:jordan} bounds, with explicit
constants, the contributions confined to a single Jordan block and separates
these resonant factors from the excess degree produced by the Jordan
structure. Theorem~\ref{thm:uniform} bounds every perturbative coefficient
uniformly in time through a comparison matrix built from unperturbed data,
for any partition of the spectrum into clusters, that is, groups of
eigenvalues treated as a single spectral sector. It controls the tail of the
series when the perturbation strength is small compared with the inverse
norm of that matrix, and it requires only a unique stationary state and a
spectral gap, so it holds for any relaxing finite-dimensional Lindbladian. Theorem~\ref{thm:EP} shows that the constants are
continuous through an exceptional point when the coalescing eigenvalues are
kept in one cluster, so that the estimates are also uniform in the distance
from it. The same comparison matrix controls the frequency-domain response
and time-dependent perturbations.
\end{revblock}

Section~\ref{sec:preliminaries} introduces Markovian open quantum systems and
fixes the notation. Section~\ref{sec:env_perturbations} derives the
perturbative corrections for time-independent and time-dependent
Liouvillians. Section~\ref{sec:exceptional_points} derives the scaling laws of
Proposition~\ref{prop:jordan}, and Sec.~\ref{sec:error_bounds} proves
Theorems~\ref{thm:uniform} and~\ref{thm:EP} and treats the frequency domain
and time-dependent perturbations. Section~\ref{sec:numerics} benchmarks the
estimates in a driven two-level system with a Liouvillian exceptional point,
including symmetry-breaking perturbations and a sweep through the exceptional
point, and in an amplitude-damping model. Section~\ref{sec:conclusion}
summarizes the results and discusses extensions and applications.
 
\section{\label{sec:preliminaries} Open quantum systems}

In this section, we fix the notation used throughout the paper and review the necessary facts about open quantum systems and their evolution.

\subsection{Open quantum evolutions}

An open quantum system is a quantum system coupled to external degrees of
freedom, usually referred to as an environment. If the system and environment are initially uncorrelated, the reduced evolution of the system is described
by a completely positive and trace-preserving map,
\begin{equation}
    \Lambda \colon \mathcal{B}(\mathcal{H})
    \to
    \mathcal{B}(\mathcal{H}),
\end{equation}
where \(\mathcal{B}(\mathcal{H})\) denotes the space of bounded linear
operators on the system Hilbert space \(\mathcal{H}\)~\cite{Breuer_2002,Davies_1974}.

In the Markovian case, the continuous-time reduced dynamics is generated by a
Gorini-Kossakowski-Sudarshan-Lindblad generator,
\begin{equation}
    \frac{d\rho}{dt}
    =
    \mathcal L[\rho]
    =
    -i[H,\rho]
    +
    \sum_{k}
    \gamma_k
    \left(
        L_k \rho L_k^\dagger
        -
        \frac{1}{2}
        \left\{
            L_k^\dagger L_k,\rho
        \right\}
    \right),
\end{equation}
where \(H=H^\dagger\) is the system Hamiltonian, \(\gamma_k\geq 0\) are
dissipative rates, and \(L_k\) are jump operators~\cite{Gorini_1976,Lindblad_1976,Davies_1974,Breuer_2002}.
The superoperator \(\mathcal L\) is called the Lindblad generator, Lindbladian or
quantum Liouvillian. In general, for a continuous-time Markovian CPTP evolution, we have a master equation
\begin{equation}
    \frac{d\rho}{dt} = \mathcal{L}(t)\rho(t),
\end{equation}
where $\mathcal{L}$ is a linear superoperator that can, in principle, be time-dependent. In this work, we use time-independent Lindbladians unless explicitly stated otherwise~\cite{Breuer_2002}.

\subsection{Eigenmode decomposition}
For a given master equation
\[
\frac{d\rho}{dt} = \mathcal L \rho(t),
\]
we can analyze properties of the formal solution $\rho(t) = e^{\mathcal L t}\rho(0)$ using the Jordan-block decomposition of $\mathcal L$~\cite{Kato_1995}. The spectral decomposition of the generator, together with the associated Liouvillian gap, provides the standard framework for the relaxation and steady-state properties of Markovian dynamics~\cite{Minganti_2018}. To this end, we first note that, since the evolution is contractive, any $\lambda \in \sigma(\mathcal L)$ satisfies $\operatorname{Re}(\lambda) \leq 0$, where $\sigma(\mathcal L)$ is the spectrum of the operator $\mathcal L$. The spectrum can then be divided into two parts: the peripheral eigenvalues, which have vanishing real part, and the nonperipheral eigenvalues, which have strictly negative real part. Let $\sigma_{\mathrm P}(\mathcal L)$ denote the peripheral spectrum of $\mathcal L$. We can decompose $\rho$ in the generalized eigenbasis associated with the Jordan canonical form of $\mathcal L$. Therefore, to obtain the time-dependent contributions, it is enough to analyze the projections of $\rho$ onto each Jordan block. If $J_m(\lambda)$ is a Jordan block of size $m$, then
\[
J_m(\lambda)=\lambda I_m+N_m,
\]
where $N_m$ is the standard nilpotent block, with $N_m^m=0$. Hence,
\begin{equation}\label{eq:jordan_exp}
    e^{J_m(\lambda)t}
    =
    e^{\lambda t}
    \sum_{k=0}^{m-1}\frac{t^k}{k!}N_m^k .
\end{equation}

First, suppose that $\lambda \in \sigma_{\mathrm P}(\mathcal L)$, so that $\operatorname{Re}(\lambda)=0$. In this case, the exponential prefactor $e^{\lambda t}$ is purely oscillatory. If $m>1$, the polynomial terms generated by the nilpotent part $N_m$ lead, in general, to unbounded growth in time. This is incompatible with the boundedness of a CPTP evolution. Therefore, Jordan blocks associated with peripheral eigenvalues must have size $m=1$. Equivalently, the peripheral spectrum of $\mathcal L$ is semisimple~\cite{Davies_1974,Breuer_2002}.

Now, let $J_m(\lambda)$ be a Jordan block associated with a nonperipheral eigenvalue,
\[
\lambda \in \sigma(\mathcal L)\setminus\sigma_{\mathrm P}(\mathcal L).
\]
Then we can write
\[
\lambda=-\gamma+i\omega,
\qquad \gamma>0.
\]
The corresponding contribution to the dynamics has the form
\begin{equation}
    \rho^{J_m(\lambda)}(t)
    =
    e^{-\gamma t}e^{i\omega t}
    \sum_{k=0}^{m-1}\frac{t^k}{k!}N_m^k
    \rho^{J_m(\lambda)}(0).
\end{equation}
Thus, every nonperipheral Jordan-block contribution is a polynomial in time multiplied by a decaying exponential. In particular,
\begin{equation}
    \rho^{J_m(\lambda)}(t)\longrightarrow 0
    \qquad \text{as } t\to\infty .
\end{equation}
The low-lying nonperipheral modes, separated from the peripheral spectrum by the Liouvillian gap, are precisely those that control the long-lived transients and metastable regimes of the dynamics~\cite{Macieszczak_2016}. Throughout this paper, $\|\cdot\|$ denotes the Hilbert--Schmidt norm on
operators, $\|X\| = (\operatorname{Tr} X^\dagger X)^{1/2}$, and superoperators
carry the norm induced by it, i.e.\ the spectral norm of their matrix
representation. With this choice $\|\mathcal{S}[X]\| \le \|\mathcal{S}\|\,\|X\|$
and $\|\mathcal{S}\mathcal{T}\| \le \|\mathcal{S}\|\,\|\mathcal{T}\|$ hold for
all superoperators $\mathcal{S},\mathcal{T}$ and operators $X$, so the
path-by-path estimates below are exact inequalities.

\section{\label{sec:env_perturbations}Perturbation theory}

Here we give a compact derivation for density-matrix perturbative corrections
and extend the construction to Liouvillians with higher-order dependence on the
perturbation parameter. We also discuss trace preservation and positivity at finite
truncation order~\cite{Yi_2000,li2014,VillegasMartinez2016PerturbationMasterEquation,Shishkov_2020}.

\subsection{Time-independent perturbation theory}

Let the perturbed Lindbladian be written as
\begin{equation}
    \mathcal{L}_{\varepsilon}
    =
    \sum_{j=0}^{p}
    \varepsilon^j\mathcal{L}_j ,
\end{equation}
where \(\mathcal{L}_0\) generates the unperturbed dynamics and
\(\varepsilon\) controls the perturbation. We assume that the propagator
\(e^{\mathcal{L}_0t}\) is known, either explicitly through a Jordan decomposition or through another convenient representation. The density matrix
is expanded as
\begin{equation}
    \rho(t)
    =
    \sum_{k\geq 0}
    \varepsilon^k\rho_k(t).
\end{equation}
Substitution into the master equation,
\begin{equation}
    \frac{d\rho}{dt}
    =
    \mathcal{L}[\rho],
\end{equation}
and comparison of equal powers of \(\varepsilon\) give
\begin{equation}\label{eqn:differential_equation_perturbation}
    \frac{d\rho_k}{dt}
    =
    \sum_{j=0}^{k}
    \mathcal{L}_j[\rho_{k-j}(t)],
    \qquad
    k\geq 0,
\end{equation}
where \(\mathcal{L}_j=0\) for \(j>p\). Equivalently,
\begin{equation}
    \frac{d\rho_k}{dt}
    =
    \mathcal{L}_0[\rho_k(t)]
    +
    \sum_{j=1}^{k}
    \mathcal{L}_j[\rho_{k-j}(t)].
\end{equation}
The zeroth-order term is therefore
\begin{equation}
    \rho_0(t)
    =
    e^{\mathcal{L}_0t}[\rho_0(0)].
\end{equation}
For \(k\geq1\), Duhamel's formula yields the recursive solution~\cite{Kato_1995}
\begin{equation}\label{eqn:general_perturbation}
\begin{split}
    \rho_k(t)
    =
    &e^{\mathcal{L}_0t}[\rho_k(0)]
    \\
    &+
    \int_0^t
    e^{\mathcal{L}_0(t-s)}
    \left[
        \sum_{j=1}^{k}
        \mathcal{L}_j[\rho_{k-j}(s)]
    \right]ds .
\end{split}
\end{equation}
Thus each correction is obtained from lower-order terms only.

A particularly important case is a linear perturbation~\cite{Yi_2000,li2014,VillegasMartinez2016PerturbationMasterEquation,Shishkov_2020},
\begin{equation}
    \mathcal{L}_{\varepsilon}
    =
    \mathcal{L}_0
    +
    \varepsilon\mathcal{L}_1 .
\end{equation}
Then
\begin{equation}
\begin{split}
    \rho_k(t)
    =
    &e^{\mathcal{L}_0t}[\rho_k(0)]
    \\
    &+
    \int_0^t
    e^{\mathcal{L}_0(t-s)}
    \mathcal{L}_1[\rho_{k-1}(s)]\,ds .
\end{split}
\end{equation}
For vanishing higher-order initial corrections,
\(\rho_k(0)=0\) for \(k\geq1\), the first two terms are
\begin{equation}\label{eq:rho1}
    \rho_1(t)
    =
    \int_0^t
    e^{\mathcal{L}_0(t-s)}
    \mathcal{L}_1
    e^{\mathcal{L}_0s}[\rho_0(0)]\,ds ,
\end{equation}
and
\begin{equation}
\begin{split}
    \rho_2(t)
    =
    \int_0^t ds
    \int_0^s du\,
    &e^{\mathcal{L}_0(t-s)}
    \mathcal{L}_1
    e^{\mathcal{L}_0(s-u)}
    \\
    &\times
    \mathcal{L}_1
    e^{\mathcal{L}_0u}[\rho_0(0)] .
\end{split}
\end{equation}
These expressions are the Liouville-space analogue of the Dyson expansion, but
written recursively in a form adapted to dissipative dynamics~\cite{Kato_1995,Yi_2000,li2014}.

\rev{As is standard for the Dyson expansion of perturbed
semigroups~\cite{EngelNagel2000,Brendle2001}, no diagonalization of the
perturbed generator is required.} If \(\mathcal{L}_0\) contains Jordan blocks, their
polynomial contributions are contained explicitly in
\(e^{\mathcal{L}_0t}\). The perturbative calculation is then reduced to
successive applications of the perturbing superoperators and time integration.

\subsection{Time-dependent perturbation theory}

The same construction applies when the Lindbladian is explicitly
time dependent~\cite{Breuer_2002,Kato_1995}. Let
\begin{equation}
    \mathcal{L}_{\varepsilon}(t)
    =
    \sum_{j = 0}^{p}
    \varepsilon^j\mathcal{L}_j(t).
\end{equation}
Expanding
\begin{equation}
    \rho(t)
    =
    \sum_{k\geq 0}
    \varepsilon^k\rho_k(t),
\end{equation}
one obtains
\begin{equation}
    \frac{d\rho_k}{dt}
    =
    \sum_{j=0}^{k}
    \mathcal{L}_j(t)[\rho_{k-j}(t)],
\end{equation}
or, equivalently,
\begin{equation}
    \frac{d\rho_k}{dt}
    =
    \mathcal{L}_0(t)[\rho_k(t)]
    +
    \sum_{j=1}^{k}
    \mathcal{L}_j(t)[\rho_{k-j}(t)].
\end{equation}
Let \(\Delta_0(t,s)\) be the propagator generated by
\(\mathcal{L}_0(t)\),
\begin{equation}
    \Delta_0(t,s)
    =
    \mathcal{T}
    \exp
    \left[
        \int_s^t
        \mathcal{L}_0(\tau)d\tau
    \right],
\end{equation}
where \(\mathcal{T}\) denotes time ordering~\cite{Breuer_2002}. The recursive solution is then
\begin{equation}
\begin{split}
    \rho_k(t)
    =
    &\Delta_0(t,0)[\rho_k(0)]
    \\
    &+
    \int_0^t
    \Delta_0(t,s)
    \left[
        \sum_{j=1}^{k}
        \mathcal{L}_j(s)[\rho_{k-j}(s)]
    \right]ds .
\end{split}
\end{equation}
For a linear perturbation,
\begin{equation}
    \mathcal{L}_{\varepsilon}(t)
    =
    \mathcal{L}_0(t)
    +
    \varepsilon\mathcal{L}_1(t),
\end{equation}
this reduces to
\begin{equation}
\begin{split}
    \rho_k(t)
    =
    &\Delta_0(t,0)[\rho_k(0)]
    \\
    &+
    \int_0^t
    \Delta_0(t,s)
    \mathcal{L}_1(s)[\rho_{k-1}(s)]\,ds .
\end{split}
\end{equation}
In particular, for \(\rho_k(0)=0\), \(k\geq1\),
\begin{equation}
    \rho_1(t)
    =
    \int_0^t
    \Delta_0(t,s)
    \mathcal{L}_1(s)
    \Delta_0(s,0)[\rho_0(0)]\,ds .
\end{equation}
Higher orders are obtained by iterating the same formula. This gives a
time-domain perturbative expansion for perturbations in open systems without requiring
instantaneous eigenvectors of the Liouvillian~\cite{Nakajima_1958,Zwanzig_1960,Fleming_2011}.

\subsection{Trace preservation and positivity}
If $\mathcal{L}$ is a Liouvillian, the trace is preserved order by order. Since the perturbed generator is trace
preserving,
\begin{equation}
    \Tr[\mathcal L[X]]=0
\end{equation}
for every operator \(X\), and since
\begin{equation}
    \mathcal L_{\varepsilon}
    =
    \sum_{j = 0}^{p}\varepsilon^j\mathcal L_j,
\end{equation}
one has
\begin{equation}
    \Tr[\mathcal L_j[X]]=0,
    \qquad
    j\geq0.
\end{equation}
Taking the trace of the recursion
\begin{equation}
    \dot\rho_k(t)
    =
    \sum_{j=0}^{k}
    \mathcal L_j[\rho_{k-j}(t)]
\end{equation}
therefore gives
\begin{equation}
    \frac{d}{dt}\Tr[\rho_k(t)]=0.
\end{equation}
For an unperturbed initial condition,
\begin{equation}
    \rho_0(0)=\rho(0),
    \qquad
    \rho_k(0)=0,
    \quad k\geq1,
\end{equation}
it follows that
\begin{equation}
    \Tr[\rho_0(t)]=1,
    \qquad
    \Tr[\rho_k(t)]=0,
    \quad k\geq1.
\end{equation}
Hence the perturbative corrections are traceless, and the expansion preserves
normalization order by order.

The same statement cannot be made for positivity at finite truncation order~\cite{Breuer_2002,Fleming_2011}.
If \(\mathcal L\) is a Lindblad generator, the exact state
\(\rho_\varepsilon(t)\) is positive. A truncated series
\begin{equation}
    \rho_\varepsilon^{[n]}(t)
    =
    \sum_{k=0}^{n}
    \varepsilon^k\rho_k(t)
\end{equation}
is instead only an approximation and need not be positive for
arbitrary \(t\) or \(\varepsilon\). Positivity is guaranteed for the full
Lindblad evolution, while the truncated expansion is valid in the
perturbative regime.

\section{Scaling and stability of perturbations}
\label{sec:exceptional_points}

In this section, we derive scaling laws for observable corrections associated
with fixed Jordan sectors of the unperturbed Lindbladian and show that under
standard relaxation hypotheses,  non-peripheral contributions decay in time.
\rev{We first note that polynomial-in-time factors appear generically, through
resonances between repeated eigenvalues along a perturbative path, and then quantify the excess degree added by Jordan
blocks.}

\subsection{\texorpdfstring{\rev{Resonant polynomial factors}}{Resonant polynomial factors}}
\label{sec:resonance}
\begin{revblock}
The \(n\)-th Dyson term is a sum over paths through the spectral sectors of
\(\mathcal L_0\). Consider a path that visits simple eigenvalues
\(\lambda_{a_0},\ldots,\lambda_{a_n}\), one per propagator, on the simplex
\(\Delta_n(t)=\{0\leq s_n\leq\cdots\leq s_1\leq t\}\). The scalar time integral
along the path is
\begin{equation}\label{eq:divided_difference}
\begin{split}
    \int_{\Delta_n(t)}
    &e^{\lambda_{a_0}(t-s_1)}e^{\lambda_{a_1}(s_1-s_2)}\cdots
    e^{\lambda_{a_n}s_n}\,d^ns.
\end{split}
\end{equation}
When the eigenvalues coincide,
\begin{equation}
\int_{\Delta_n(t)}
    e^{\lambda t}\,d^ns = e^{\lambda t}\,\operatorname{Vol}(\Delta_n(t)) = \frac{t^n}{n!}\,e^{\lambda t}.
\end{equation}
Polynomial factors in time therefore appear already for a diagonalizable
generator. They are resonances, produced by repeated or nearly repeated
eigenvalues along a path, and they are not specific to exceptional points. A
Jordan block produces the same coalescence of exponents inside a single
propagator, and this is what adds the excess degree quantified below.
\end{revblock}

\subsection{\label{sec:jordan}Polynomial time dependence from Jordan blocks}

The advantage of the time-domain formulation becomes especially transparent
when the unperturbed Lindbladian is not diagonalizable~\cite{Kato_1995,Heiss_2012,Ashida_2020}.
\rev{By Eq.~\eqref{eq:jordan_exp}, a Jordan block
\(J_m(\lambda)=\lambda I_m+N_m\) of size \(m\) produces polynomial-in-time
factors up to degree \(m-1\), and the highest possible polynomial contribution
inside the generalized eigenspace is proportional to} $t^{m-1}e^{\lambda t}$. Whether this highest-degree term appears in a given trajectory or observable
depends on the projection of the initial state and of the observable onto the
corresponding generalized eigenvectors. \rev{At an exceptional point, the norm
\(\|N_m^{m-1}P_J\|\) of the top nilpotent power, with \(P_J\) the spectral
projection onto the block, is the spectral response
strength~\cite{Wiersig_2022,BidSchomerus2025}, which sets the eigenvalue splitting under
perturbation.}

This structure is inherited by the perturbative recursion. \rev{For a
polynomial perturbation \(\mathcal L_\varepsilon=\mathcal L_0+\varepsilon\mathcal
L_1+\cdots+\varepsilon^p\mathcal L_p\), the \(k\)-th correction of
Eq.~\eqref{eqn:general_perturbation} is a
sum over compositions
\(k=j_1+\cdots+j_n\), \(1\leq j_a\leq p\), each with \(n\) insertions integrated
over the simplex \(\Delta_n(t)\):}
\begin{equation}
\begin{split}
    \rho_{k, \mathcal J}(t)
    =
    \int_{\Delta_n(t)}
    &
    e^{\mathcal L_0(t-s_1)}
    \mathcal L_{j_1}
    e^{\mathcal L_0(s_1-s_2)}
    \\
    &\times
    \mathcal L_{j_2}
    \cdots
    \mathcal L_{j_n}
    e^{\mathcal L_0s_n}
    [\rho(0)]
    \,d^ns ,
\end{split}
\label{eq:dyson_k_several}
\end{equation}
\rev{where \(\mathcal{J} = (j_1,\dots,j_n)\) denotes a fixed composition.}
\rev{In the contribution in which all \(n+1\)
propagators are projected onto the block \(J_m(\lambda)\)~\cite{Kato_1995}, each
propagator carries a polynomial of degree at most \(m-1\) and the integration a
factor \(t^n\), so the maximal degree is \((n+1)m-1\): \(t^{2m-1}e^{\lambda t}\)
at first order and \(t^{(k+1)m-1}e^{\lambda t}\) at order \(k\). This is an
upper scaling, and lower powers occur when the initial state, the perturbation, or
the observable has no component on the highest generalized eigenvectors. For
mixed paths, the exponentials of the visited blocks combine into Eq.~\eqref{eq:divided_difference}. The factor \(t^n\) is the
resonance of Sec.~\ref{sec:resonance} and is present already for \(m=1\). Therefore, the excess \((n+1)(m-1)\) is produced by the Jordan structure. To make this
quantitative, note that by Eq.~\eqref{eq:jordan_exp} each projected propagator
satisfies \(\|e^{\mathcal L_0\tau}P_J\|\leq
e^{-\gamma\tau}\sum_{r}\kappa_r\tau^r/r!\), with \(\gamma=-\operatorname{Re}\lambda\)
and \(\kappa_r=\|N_m^rP_J\|\). Bounding the \(n+1\) propagators in this way,
using \(e^{-\gamma(\tau_0+\cdots+\tau_n)}=e^{-\gamma t}\) with the time gaps
\(\tau_0=t-s_1, \tau_1 = s_1-s_2,\ldots,\tau_n=s_n\), and integrating with the Dirichlet formula}
\[
\int_{\sum_i\tau_i=t}\prod_{i=0}^{n}\frac{\tau_i^{r_i}}{r_i!}
d\tau_1\cdots d\tau_n=\frac{t^{n+|r|}}{(n+|r|)!},
\] 
\rev{with \(|r|:=\sum_ir_i\), gives an
explicit envelope.}
This discussion leads us to the \rev{following proposition}:
\begin{revproposition}[Single-block Jordan scaling of perturbative corrections]\label{prop:jordan}
Let
\[
    \mathcal L_\varepsilon
    =
    \mathcal L_0+\varepsilon\mathcal L_1
    +\cdots+\varepsilon^p\mathcal L_p
\]
be a finite-dimensional perturbed Lindbladian, and let \(J_m(\lambda)\) be a
Jordan block of \(\mathcal L_0\) of size \(m\). For a fixed composition
\(\mathcal J=(j_1,\ldots,j_n)\) contributing to the \(k\)-th perturbative
correction, with
\[
    k=j_1+\cdots+j_n,
    \qquad
    1\leq j_a\leq p,
\]
the single-block contribution in which all unperturbed propagators are
projected onto \(J_m(\lambda)\) satisfies \rev{the envelope}
\begin{equation}\label{eq:envelope}
    \rev{\big\|\rho^{(J)}_{k,\mathcal J}(t)\big\|
    \leq
    \beta_{\mathcal J}\,e^{-\gamma t}
    \!\!\sum_{r_0,\ldots,r_n=0}^{m-1}
    \Big(\prod_{i=0}^{n}\kappa_{r_i}\Big)
    \frac{t^{\,n+|r|}}{(n+|r|)!},}
\end{equation}
\rev{with \(\beta_{\mathcal J}:= ( \prod_{i=1}^{n}\|P_J\mathcal L_{j_i}P_J\| ) \,
\|P_J\rho(0)\|\), and hence} the upper scaling
\begin{equation}
    \rho_{k,\mathcal J}^{(J)}(t)
    =
    O\!\left(
        t^{(n+1)m-1}e^{\lambda t}
    \right).
\end{equation}
Consequently, the largest possible single-block Jordan-polynomial scaling at
order \(k\) is
\begin{equation}
    \rho_k^{(J)}(t)
    =
    O\!\left(
        t^{(k+1)m-1}e^{\lambda t}
    \right),
\end{equation}
provided that the corresponding highest-insertion contribution is nonvanishing.
\rev{Of the maximal degree \((n+1)m-1\), the part \(n\) is a resonance present
for any block, and the excess \((n+1)(m-1)\) is produced by the Jordan
structure.}
\end{revproposition}
\rev{For a mixed path through Jordan blocks \(a_0,\ldots,a_n\), of sizes
\(m_{a_i}\) and decay rates \(\gamma_{a_i}\), the same argument gives
Eq.~\eqref{eq:envelope} with \(\gamma\) replaced by \(\min_i\gamma_{a_i}\),
\(\kappa_{r_i}\) by \(\|N_{a_i}^{r_i}P_{a_i}\|\) with \(r_i\le m_{a_i}-1\), and
\(\beta_{\mathcal J}\) by
\(\big(\prod_{i=1}^n\|P_{a_{i-1}}\mathcal L_{j_i}P_{a_i}\|\big)\|P_{a_n}\rho(0)\|\).
The maximal degree is therefore \(\sum_i m_{a_i}-1\le(n+1)m_{\max}-1\), so no
path with \(n\) insertions has a higher degree in \(t\) than the single-block
contribution of a block of maximal size. This concerns the degree only: a mixed
path can still dominate in magnitude, for instance when it visits a more slowly
decaying block.}

\subsection{Transient amplification and stability of observable responses}
\label{sec:transient_amplification}

We now address the following question: how do perturbations affect observables? \rev{By the estimates above, the
\(k\)-th order single-block contribution to the expectation value of an
observable \(\mathcal O\),
\(\delta\langle\mathcal O\rangle_k^{(J)}(t):=\operatorname{Tr}[\mathcal
O\rho_k^{(J)}(t)]\), satisfies}
\begin{equation}
    \delta\langle \mathcal O\rangle_k^{(J)}(t)
    =
    O\!\left(
        t^{(k+1)m-1}e^{-\gamma t}
    \right).
\end{equation}
Thus, although Jordan blocks can enhance the transient response by polynomial
factors in time, these contributions are exponentially damped whenever
\(\gamma>0\). This polynomial transient growth followed by asymptotic decay is
the hallmark of the non-normality of the generator, and is captured
systematically by its pseudospectrum~\cite{Trefethen_Embree_2005}. At a
Liouvillian exceptional point the defectiveness is maximal and the associated
response strength is correspondingly enhanced~\cite{Wiersig_2022,Hashemi_2022}.

\begin{revblock}
Boundedness and decay of a contribution bounded by \(C\,t^re^{-\gamma t}\)
follow from \(\gamma>0\) alone, and the bound peaks at \(t_*=r/\gamma\),
with \(\sup_tt^re^{-\gamma t}/r!\leq(2\pi r)^{-1/2}\gamma^{-r}\), so as
\(\gamma\) decreases the amplification grows and occurs later, while the
contribution still vanishes asymptotically. The exceptional point enters only
quantitatively. It raises the maximal degree from \(n\) to \((n+1)m-1\), so the
peak scales as \(\gamma^{-(n+1)m+1}\) instead of \(\gamma^{-n}\) and occurs
later. Integrated over time,
\(\int_0^\infty e^{-\gamma t}t^r/r!\,dt=\gamma^{-r-1}\); for the propagator of a
defective block this raises the inverse power of the gap from \(1\) to \(m\),
which is how defectiveness enters the uniform bounds of
Sec.~\ref{sec:error_bounds}. 

Whether the terms produced by the Jordan structure are visible is set by the
dimensionless ratio \(\eta:=\kappa_{m-1}/(\kappa_0\gamma^{m-1})\). Up to the
factor \(1/(m-1)!\), it is the ratio between the highest-degree term and the
purely exponential term of Eq.~\eqref{eq:jordan_exp} at the lifetime
\(t=1/\gamma\) of the mode, and the power of \(\gamma\) makes it independent of
units.
\end{revblock}

\section{\label{sec:error_bounds} Uniform-in-time bounds for finite truncations}
We show that under suitable relaxing assumptions, the perturbative coefficient series is
uniformly controlled in time. Here, any consistent
submultiplicative norm leads to the same conclusions. \rev{The constants
defined in the bound depend on this choice.}
\subsection{\texorpdfstring{\rev{Comparison-matrix bound}}{Comparison-matrix bound}}
\label{sec:comparison}
\begin{revblock}
The idea of the proof is to insert the resolution of the identity given by the
spectral projectors of \(\mathcal L_0\) between the propagators of each term of
the Dyson expansion of the \(k\)-th order correction. Each term then becomes a
sum over sequences of projectors, which we call paths. In each path, we bound the inner propagators by
their time integrals and the last one by its supremum. Summing over all paths that give nonzero contributions turns the bound into a power of a nonnegative matrix, the comparison matrix \(A\).
\end{revblock}

First, consider a linear perturbation
\(\mathcal L_\varepsilon=\mathcal L_0+\varepsilon\mathcal L_1\). Writing
\(\rho(t)=\sum_{k=0}^{\infty}\varepsilon^k\rho_k(t)\), the \(k\)-th coefficient, for \(k\geq 1\), is
\begin{equation}
\begin{split}
    \rho_k(t)
    =
    \int_{\Delta_k(t)}
    &
    e^{\mathcal L_0(t-s_1)}
    \mathcal L_1
    e^{\mathcal L_0(s_1-s_2)}
    \\
    &\times
    \mathcal L_1
    \cdots
    \mathcal L_1
    e^{\mathcal L_0s_k}
    [\rho(0)]
    \,d^ks .
\end{split}
\label{eq:dyson_k}
\end{equation}
We assume that the only spectral subspace of \(\mathcal L_0\) with zero real
part is the one-dimensional stationary subspace, that is, $\mL_0$ has a unique stationary state and no other eigenvalue on the imaginary axis. Let \(P_0\) be the
corresponding spectral projection. Then
\(P_0=|\rho_{\mathrm{ss}}\rangle\langle I |\)\rev{, where \(|X\rangle\) denotes an
operator \(X\) regarded as a vector of Liouville space, with the
Hilbert-Schmidt inner product \(\langle X|Y\rangle=\operatorname{Tr}(X^\dagger
Y)\), so that \(\langle I|X\rangle=\operatorname{Tr}X\). Here}
\(\mathcal L_0|\rho_{\mathrm{ss}}\rangle=0\) expresses stationarity and
\(\langle I|\mathcal L_0=0\) trace preservation. Since \(\mathcal L_1\) is trace preserving,
\(\langle I|\mathcal L_1=0\), and therefore
\begin{equation}
    P_0\mathcal L_1=0 .
\end{equation}
This identity removes the secular terms that would otherwise be generated by
repeated insertions of the stationary projection.

\rev{We group the nonzero eigenvalues of \(\mathcal L_0\) into disjoint sets
\(\sigma_a\), which we call clusters, labeled by a finite index set \(S\).
The label \(0\notin S\) is reserved for the stationary sector. The spectral projector of a
cluster is its Riesz projector, which can be represented as the sum of the projectors onto the
generalized eigenspaces of the eigenvalues in \(\sigma_a\). These projectors
satisfy \(P_aP_b=\delta_{ab}P_a\), \(P_0+\sum_{a\in S}P_a=I\), and
\([\mathcal L_0,P_a]=0\). A cluster may contain a single eigenvalue, in which
case \(P_a\) projects onto its Jordan sector, or several. The finest partition
takes one eigenvalue per cluster, and the coarsest puts all nonzero eigenvalues
in a single cluster, with \(P_a=I-P_0\). Near an exceptional point, we keep the
coalescing eigenvalues in the same cluster (Sec.~\ref{sec:uniform_EP}). For a
Jordan sector \(a\in S\),} write
\(\mathcal L_0P_a=(\lambda_a I+N_a)P_a\), with
\(\lambda_a=-\gamma_a+i\alpha_a\), \(\gamma_a>0\), and \(N_a^{n_a}=0\). Then
\begin{equation}
    e^{\mathcal L_0s}P_a
    =
    e^{\lambda_as}
    \sum_{r=0}^{n_a-1}
    \frac{s^r}{r!}N_a^rP_a .
\end{equation}
Hence
\begin{equation}
    \left\|e^{\mathcal L_0s}P_a\right\|
    \leq
    e^{-\gamma_as}K_a(s),
\end{equation}
where
\(K_a(s):=\sum_{r=0}^{n_a-1}s^r\|N_a^rP_a\|/r!\). The corresponding integrated weight is
\begin{equation}\label{eq:Ha}
\begin{split}
    H_a
    :=
    \rev{\int_0^\infty \left\|e^{\mathcal L_0s}P_a\right\| ds}
    &\mathrel{\rev{\leq}}
    \int_0^\infty e^{-\gamma_as}K_a(s)\,ds
    \\
    &=
    \sum_{r=0}^{n_a-1}
    \frac{\left\|N_a^rP_a\right\|}
         {\gamma_a^{r+1}} .
\end{split}
\end{equation}
\rev{The weight \(H_a\) is the integrated response of sector \(a\): the total
weight that the sector propagates over all times. Only an
upper bound on \(H_a\) is needed below, so the closed form on the right may be
used in place of the integral. For a cluster that contains several eigenvalues, we use the integral itself.}
We also define the supremum weight
\begin{equation}
    M_a
    :=
    \sup_{s\geq 0}
    \left\|e^{\mathcal L_0s}P_a\right\|,
    \qquad
    a\in S ,
\end{equation}
\rev{which represents the largest transient amplification inside sector \(a\).}
For the stationary projection, we use
\(e^{\mathcal L_0s}P_0=P_0\) directly.

To estimate the simplex integrals, introduce the time gaps
\(\tau_0=t-s_1\), \(\tau_1=s_1-s_2\), \ldots, \(\tau_k=s_k\). Then
\(\tau_i\geq 0\) and \(\sum_{i=0}^{k}\tau_i=t\). After inserting the
resolution of the identity \(I=P_0+\sum_{a\in S}P_a\), each term is a sum over
block paths \((a_0,a_1,\ldots,a_k)\). The identity \(P_0\mathcal L_1=0\) implies that every surviving path has
\(a_i\neq 0\) for \(i<k\). Thus, the stationary projection can appear only at
the rightmost end of a surviving path.

For a fixed surviving path, the first \(k\) propagator intervals are bounded by
their integrated weights \(H_a\), while the rightmost interval is bounded by a
supremum weight. This gives a time-uniform estimate for each perturbative
coefficient.

Define the comparison matrix
\begin{equation}
    A_{ab}
    :=
    H_a
    \left\|P_a\mathcal L_1P_b\right\|,
    \qquad
    a\in S,\quad b\in S\cup\{0\},
\end{equation}
and set \(A_{0b}=0\). Define also the initial weights
\begin{equation}
    y_b
    :=
    \begin{cases}
        M_b\left\|P_b\rho(0)\right\|,
        & b\in S,\\[4pt]
        \left\|P_0\rho(0)\right\|,
        & b=0.
    \end{cases}
\end{equation}
Then the simplex estimate gives, for every \(k\geq 1\),
\begin{equation}\label{eq:path_bound}
    \sup_{t\geq 0}\|\rho_k(t)\|
    \leq
    \left\|A^k y\right\|_1 .
\end{equation}
\rev{The matrix \(A\) is the weighted adjacency matrix of the directed graph of
spectral sectors, with an edge \(b\to a\) when \(P_a\mathcal L_1P_b\neq0\). The
entry \(A_{ab}\) describes one transfer from sector \(b\) to sector \(a\),
multiplied by the integrated response \(H_a\) of the receiving sector, and
\((A^k)_{ab}\) sums all \(k\)-step paths. The vector \(y\) carries the initial
weight of each sector, including its transient amplification \(M_b\). Since all
entries are nonnegative, \(\|A^ky\|_1=\mathbf 1^{\mathsf T}A^ky\).}
Therefore, if
\begin{equation}
    R:=\|A\|_1,
    \qquad
    C_\rho:=\|y\|_1,
\end{equation}
\rev{where \(\|A\|_1=\max_b\sum_aA_{ab}\) is the maximal column sum,}
then
\begin{equation}
    \sup_{t\geq 0}\|\rho_k(t)\|
    \leq
    C_\rho R^k .
\end{equation}
The relevant small parameter is therefore not \(\varepsilon\) itself but
\begin{equation}
    q:=|\varepsilon|R .
\end{equation}
If
\begin{equation}
    q<1,
\end{equation}
then the perturbative coefficient series is uniformly summable in time. The
tail after order \(N\) satisfies
\begin{equation}
\begin{split}
    \sup_{t\geq 0}
    \Bigg\|
        \sum_{k=N+1}^{\infty}
        \varepsilon^k\rho_k(t)
    \Bigg\|
    &\leq
    \sum_{k=N+1}^{\infty}
    |\varepsilon|^k
    \sup_{t\geq 0}\|\rho_k(t)\|
    \\
    &\leq
    C_\rho
    \sum_{k=N+1}^{\infty}q^k
    \\
    &=
    C_\rho
    \frac{q^{N+1}}{1-q}.
\end{split}
\end{equation}
In particular, to guarantee that the total correction beyond zeroth order is
bounded by a prescribed tolerance \(\zeta>0\), it is sufficient that
\rev{\(q\leq\zeta/(C_\rho+\zeta)\).}
Thus, a large value of \(R\) does not by itself invalidate perturbation theory;
it only makes explicit that the effective expansion parameter is
\(|\varepsilon|R\). The weights \(H_a\), which encode the spectral gaps
\(\gamma_a\) and the nilpotent structure \(N_a\) of each block, play here the
role that the Drazin inverse of the generator plays in the geometric response
theory of Lindbladians~\cite{Albert_2016}.

\begin{theorem}[Uniform coefficient estimate for linear perturbations] \label{thm:uniform}
Let
\[
    \mathcal L_\varepsilon
    =
    \mathcal L_0+\varepsilon\mathcal L_1
\]
be a perturbed finite-dimensional Lindbladian \rev{(more generally, \(\mathcal
L_1\) may be any trace-preserving superoperator and \(\varepsilon\) complex)}. Assume that \(\mathcal L_0\) has
a unique stationary state, that all other spectral blocks satisfy
\[
    \operatorname{Re}\lambda<0,
\]
and that \(\mathcal L_1\) is trace preserving. Let \(A\), \(R\), and \(C_\rho\) be defined as above, for any partition of the nonzero spectrum of \(\mathcal L_0\) into clusters. If
\begin{equation}
    q:=|\varepsilon|R<1,
\end{equation}
then the perturbative coefficient series is uniformly summable in time and its
tail satisfies
\begin{equation}\label{eq:tail_q}
\begin{split}
    \sup_{t\geq 0}
    \Bigg\|
        \sum_{k=N+1}^{\infty}
        \varepsilon^k\rho_k(t)
    \Bigg\|
    \leq
    C_\rho
    \frac{q^{N+1}}{1-q}.
\end{split}
\end{equation}
\end{theorem}
Observe that stable Jordan blocks enter the constant \(R\) through the factors
\[    H_a
    \mathrel{\rev{\leq}}
    \sum_{r=0}^{n_a-1}
    \frac{\left\|N_a^rP_a\right\|}
         {\gamma_a^{r+1}},
\] 
so larger Jordan blocks and smaller spectral gaps increase the allowed
transient amplification but do not produce secular growth under the above
assumptions.

\begin{revblock}
Keeping the full matrix \(A\), rather than its norm \(R\), sharpens the
estimate. Its spectral radius satisfies \(r(A)\le\|A\|_1=R\). Transitions forbidden by
the perturbation, that is, zero entries of \(A\), lower \(r(A)\) but not \(R\). Summing
Eq.~\eqref{eq:path_bound} over \(k>N\) gives the Neumann series of
\(|\varepsilon|A\), so for \(|\varepsilon|\,r(A)<1\) the series is still
uniformly summable in time, with
\begin{equation}\label{eq:tail_refined}
    \sup_{t\geq 0}
    \Bigg\|
        \sum_{k=N+1}^{\infty}
        \varepsilon^k\rho_k(t)
    \Bigg\|
    \leq
    \mathbf 1^{\mathsf T}\big(|\varepsilon|A\big)^{N+1}
    \big(I-|\varepsilon|A\big)^{-1}y .
\end{equation}
This bound is never larger than that of Theorem~\ref{thm:uniform}, since
\(\mathbf 1^{\mathsf T}A^ky\leq\|A\|_1^k\|y\|_1\).
\end{revblock}

We stress that $q<1$ is a sufficient condition for uniform
summability, not a radius of convergence. For a finite-dimensional generator
the series $\sum_k \varepsilon^k \rho_k(t)$ converges for every $\varepsilon$ at
each fixed $t$, since $e^{\mathcal{L}_\varepsilon t}$ is entire in $\varepsilon$;
the role of $q<1$ is to make this convergence uniform on $[0,\infty)$.
\rev{With eigenmode projectors, the constants diverge as an exceptional point
is approached, because the projectors become strongly nonorthogonal. However, we can use
cluster projectors so that they remain bounded, as we show in the next subsection.}
Proposition~\ref{prop:jordan} bounds a single
block-diagonal term \rev{and describes the time profile of an individual
sector}. The full
correction $\rho_k(t)$ is a finite sum of such block-path contributions,
including mixed paths whose exponential factor combines the eigenvalues of the
blocks visited along the path. The collective control of this sum, rather than
of an individual term, is the content of Theorem~\ref{thm:uniform} \rev{and
Eq.~\eqref{eq:tail_refined}}.

\begin{revblock}
The series has an exact radius of uniform summability, which is in general
hard to compute analytically. The bound of Theorem~\ref{thm:uniform} and the
parameter \(q\), by contrast, are computed from unperturbed data alone. The long-time
limits of the coefficients,
\begin{equation}\label{eq:drazin}
    \rho_k(\infty)=\big(-\mathcal L_0^{D}\mathcal L_1\big)^k\rho_{\mathrm{ss}},
    \qquad
    \mathcal L_0^{D}=(\mathcal L_0+P_0)^{-1}-P_0,
\end{equation}
with \(\mathcal L_0^{D}\) the Drazin inverse, are the Taylor coefficients of the
exact perturbed steady state of Ref.~\cite{NagibWalker2025}. The bound of
Theorem~\ref{thm:uniform} controls, in addition, the transient part of every
coefficient, which determines finite-time quantities.
\end{revblock}

\begin{revblock}
\subsection{Uniformity in the distance from an exceptional point}
\label{sec:uniform_EP}

Theorem~\ref{thm:uniform} holds for every spectral partition, and near an
exceptional point the choice matters. Consider a continuous family
\(\mathcal L_0(\delta)\) with an exceptional point of order two at
\(\delta=0\). If the partition separates the two eigenvalues
\(\lambda_\pm=\bar\lambda\pm\mu(\delta)\) that coalesce there, the separate
projectors
\begin{equation}\label{eq:Ppm}
\begin{split}
    P_\pm(\delta)
    &=
    \pm\frac{\big(\mathcal L_0(\delta)-\lambda_\mp(\delta)\big)P_c(\delta)}
           {2\mu(\delta)},
    \\
    \|P_\pm(\delta)\|
    &=
    \frac{\|N\|}{2|\mu(\delta)|}\big(1+o(1)\big),
\end{split}
\end{equation}
diverge as \(|\delta|^{-1/2}\) at a generic exceptional point. Here
\(P_c=P_++P_-\) is the cluster projector and \(N\) is the nilpotent part at \(\delta=0\). Every constant
built from \(P_\pm\) inherits this divergence. Keeping the coalescing
eigenvalues together removes it. For a generic exceptional point of order \(m\),
the \(m\) coalescing eigenvalues split as
\(|\delta|^{1/m}\)~\cite{Seyranian_2005}, and the separate projectors
\(P_j=\prod_{k\neq j}(\mathcal L_0-\lambda_k)P_c/\prod_{k\neq j}(\lambda_j-\lambda_k)\),
whose numerators tend to \(N^{m-1}P_c\), diverge as \(|\delta|^{-(m-1)/m}\).
Theorem~\ref{thm:EP} does not depend on the order of the exceptional point, since its proof uses only contour integrals around each cluster.

\begin{theorem}[Uniformity in the distance from an exceptional point]\label{thm:EP}
Let \(\mathcal L_0(\delta)\) be a continuous family of relaxing Lindbladians,
let \(\mathcal L_1\) be trace preserving, and partition the nonzero spectrum of
\(\mathcal L_0(0)\) into clusters \(\sigma_a(0)\). For \(\delta\) near \(0\),
let \(P_a(\delta)\) be the total Riesz projector of the eigenvalues of
\(\mathcal L_0(\delta)\) that tend to \(\sigma_a(0)\). Then the constants
\(H_a\), \(M_a\), \(A\), \(R\), and \(C_\rho\) of Theorem~\ref{thm:uniform}
are continuous in \(\delta\) near \(0\), and at \(\delta=0\) they are those of
the generalized eigenspaces of \(\mathcal L_0(0)\). In particular, the bound of
Theorem~\ref{thm:uniform} is uniform in the distance from the exceptional
point.
\end{theorem}
Theorem~\ref{thm:EP} follows from the contour representation and the
continuity of the Riesz projectors~\cite{Kato_1995}, combined with a contour
bound on the propagator that is uniform in \(\delta\). Choose disjoint simple
closed curves \(\Gamma_a\subset\{\operatorname{Re}z<0\}\), each enclosing
\(\sigma_a(0)\) and no other eigenvalue of \(\mathcal L_0(0)\). Eigenvalues depend continuously on \(\delta\) and each \(\Gamma_a\) is compact,
so there is a compact neighborhood \(U'\) of \(0\) on which no eigenvalue of
\(\mathcal L_0(\delta)\) crosses any curve \(\Gamma_a\). The eigenvalues enclosed by
\(\Gamma_a\) at \(\delta=0\) remain inside it. For \(\delta\in U'\),
\(P_a(\delta)\) is the Riesz projector inside \(\Gamma_a\), and
\((z,\delta)\mapsto(z-\mathcal L_0(\delta))^{-1}\) is continuous on
\(\Gamma_a\times U'\). The Riesz projectors
\(P_a(\delta)=(2\pi i)^{-1}\oint_{\Gamma_a}(z-\mathcal L_0(\delta))^{-1}dz\) are
therefore continuous, and analytic if \(\mathcal L_0(\delta)\) is
analytic~\cite{Kato_1995}. Writing
\begin{equation}
    e^{\mathcal L_0(\delta)t}P_a(\delta)
    =
    \frac{1}{2\pi i}\oint_{\Gamma_a}e^{zt}\big(z-\mathcal L_0(\delta)\big)^{-1}dz
\end{equation}
and letting \(-\omega_a:=\max_{z\in\Gamma_a}\operatorname{Re}z<0\), \(\ell_a\) be
the length of \(\Gamma_a\), and
\(C_a:=\sup_{\Gamma_a\times U'}\|(z-\mathcal L_0(\delta))^{-1}\|\), we obtain
\(\|e^{\mathcal L_0(\delta)t}P_a(\delta)\|\leq(2\pi)^{-1}\ell_a C_a e^{-\omega_at}\)
uniformly in \(\delta\in U'\). The integrand of \(H_a(\delta)\) is therefore
continuous in \(\delta\) and dominated by an integrable function, so \(H_a\) is
continuous by dominated convergence. The same bound makes
\(\|e^{\mathcal L_0(\delta)t}P_a(\delta)\|\) uniformly small at large \(t\), and
the function is continuous in \(\delta\) uniformly on compact time intervals,
so \(M_a\) is continuous. \(A\) and \(y\) are continuous functions of these
quantities and of the projectors, and \(R\) and \(r(A)\) are continuous
functions of the entries of \(A\). Being continuous on the compact set \(U'\), all constants are bounded there,
which gives the uniformity. The statement at \(\delta=0\) is a standard
property of Riesz projectors~\cite{Kato_1995}. In particular, the cluster weight \(H_c\) is continuous through the
exceptional point, while weights built from the separate projectors
\(P_\pm\) inherit the divergence of Eq.~\eqref{eq:Ppm}. The same holds for
weights built from the individual eigenprojectors at a higher-order
exceptional point.

The constants of different partitions are not ordered: finer partitions record which sectors the perturbation connects, through the zero entries of \(A\), while coarser ones avoid ill-conditioned projectors.
Near an exceptional point, we keep each coalescing group in one cluster and
resolve the rest as finely as the conditioning allows.

The coarsest partition, \(\{P_0,\,Q=I-P_0\}\), shows the relation with the
bound of Szehr and Wolf~\cite{SzehrWolf2013}. Both start from the Duhamel
identity \(e^{\mathcal L_\varepsilon t}-e^{\mathcal L_0t}=\varepsilon\int_0^t
e^{\mathcal L_0(t-s)}\mathcal L_1e^{\mathcal L_\varepsilon s}\,ds\). Since
\(\mathcal L_1\) is trace preserving, the unperturbed propagator acts only on
traceless operators. Szehr and Wolf assume an exponential convergence bound
\(\|e^{\mathcal L_0t}-P_0\|\leq Ke^{-\nu t}\), with \(K, \nu>0\).
Since \(P_0X=0\) for traceless \(X\), this gives, in the trace norm,
\(\|e^{\mathcal L_0t}X\|\leq\min(1,Ke^{-\nu t})\|X\|\). They bound the
perturbed propagator by one, because it is a contraction, and obtain
\(\sup_t\|e^{\mathcal L_\varepsilon t}-e^{\mathcal L_0t}\|\leq|\varepsilon|\,
\|\mathcal L_1\|(\log K+1)/\nu\), with
\((\log K+1)/\nu=\int_0^\infty\min(1,Ke^{-\nu t})\,dt\). In our construction
the corresponding quantity is the weight \(H_Q=\int_0^\infty\|e^{\mathcal L_0t}Q\|\,dt\), the time integral of the same
propagator on traceless operators, onto which \(Q\) maps. The two estimates therefore have the same
structure at first order and differ in the higher orders. The contractivity
of the perturbed semigroup controls the whole deviation at once, linearly in
\(|\varepsilon|\), but requires the perturbed evolution to remain positive.
Iterating the identity instead bounds every coefficient \(\rho_k\), requires
only that \(\mathcal L_1\) be trace preserving, and holds for complex
\(\varepsilon\), which is what controls the convergence of the series. Finer
partitions then add what a single sector \(Q\) discards: the Jordan structure
of each block and the connectivity of the perturbation.

\subsection{Frequency domain}
\label{sec:frequency}

For \(\operatorname{Re}s>0\), let
\(\hat\rho_k(s):=\int_0^\infty e^{-st}\rho_k(t)\,dt\) and
\(G_0(s):=(s-\mathcal L_0)^{-1}\). The Dyson recursion is a convolution in
time, so \(\hat\rho_k(s)=G_0(s)[\mathcal L_1G_0(s)]^k\rho(0)\). Summing over
\(k\) gives, for \(|\varepsilon|\) small enough,
\begin{equation}
    \sum_k\varepsilon^k\hat\rho_k(s)=(s-\mathcal L_\varepsilon)^{-1}\rho(0),
\end{equation}
the Born series of the perturbed resolvent. Since \(0\) is the only eigenvalue
of \(\mathcal L_0\) on the imaginary axis, this expression extends
\(\hat\rho_k\) analytically to \(\operatorname{Re}s\geq0\), \(s\neq0\). There,
\(\|G_0(s)P_a\|\leq\int_0^\infty\|e^{\mathcal L_0t}P_a\|\,dt=H_a\) for
\(a\in S\), while \(G_0(s)P_0=P_0/s\) and \(P_0\mathcal L_1=0\) confine the
stationary sector to the last position of a path. Hence, for
\(\operatorname{Re}s\geq0\), \(s\neq0\),
\begin{equation}\label{eq:freq_bound}
    \|\hat\rho_k(s)\|
    \leq
    \mathbf 1^{\mathsf T}A^k\hat y(s),
\end{equation}
with \(\hat y_b=H_b\|P_b\rho(0)\|\) for \(b\in S\) and
\(\hat y_0=\|P_0\rho(0)\|/|s|\).
The time-domain weights therefore control the frequency-domain response with
the same matrix \(A\), uniformly on the closed right half-plane up to the
explicit stationary pole. For a Jordan block, the Laplace transform of
Eq.~\eqref{eq:jordan_exp} gives
\((s-\mathcal L_0)^{-1}P_J=\sum_{r=0}^{m-1}N_m^rP_J/(s-\lambda)^{r+1}\), a pole
of order \(m\) whose leading coefficient has norm \(\|N_m^{m-1}P_J\|\), the
spectral response strength~\cite{Wiersig_2022} behind the super-Lorentzian
response studied by Bid and Schomerus~\cite{BidSchomerus2025}. Hence the time-domain and frequency-domain descriptions are controlled by the same
data. In this sense the expansion is a time-domain counterpart of the regular
response theory of Ref.~\cite{BidSchomerus2025}, and, by
Eq.~\eqref{eq:freq_bound}, its bounds carry over directly to the frequency
domain.

\subsection{Time-dependent perturbations}
\label{sec:time_dependent}

The same path expansion applies to a time-dependent perturbation
\(\mathcal L_1(t)\) of a time-independent \(\mathcal L_0\), with the
time-ordered coefficients obtained from Eq.~\eqref{eq:dyson_k} by replacing the
\(i\)-th insertion with \(\mathcal L_1(s_i)\). If \(\mathcal L_1(t)\) is
piecewise continuous and trace preserving at every time, \(P_0\mathcal
L_1(t)=0\) still removes the secular paths, and with
\(b_{ab}:=\sup_t\|P_a\mathcal L_1(t)P_b\|<\infty\),
Theorem~\ref{thm:uniform} and Eq.~\eqref{eq:tail_refined} hold with
\(A_{ab}:=H_a\,b_{ab}\). For
\(\mathcal L_1(t)=f(t)\mathcal L_1\) with \(\sup|f|\leq1\), the constants are
those of the static perturbation. A time-dependent \(\mathcal L_0(t)\) is not
covered: its spectral projectors are not conserved, so our method does not
extend to it directly.
\end{revblock}

\section{\label{sec:numerics}Numerical validation}

We benchmark the estimates of Secs.~\ref{sec:exceptional_points}
and~\ref{sec:error_bounds} on a minimal Lindbladian model of the single-ion
exceptional-point setup of Ref.~\cite{Zhang_2022}, which realizes a genuine
order-two Liouvillian exceptional point and lets us test
Theorem~\ref{thm:uniform} and Proposition~\ref{prop:jordan} directly.
\rev{Section~\ref{sec:numerics-symmetry} adds two perturbations that break the
symmetry of the model, Sec.~\ref{sec:numerics-sharpness} compares the bound
with a diagonalizable model in which it is sharp, and
Sec.~\ref{sec:numerics-sweep} follows the constants through the exceptional
point, which tests Theorem~\ref{thm:EP}.}

\subsection{\label{sec:numerics-model}Model and exceptional point}

Consider a driven two-level system with a single decay channel,
\begin{equation}
  H(\Omega)=\tfrac{\Omega}{2}\,\sigma_x, \qquad
  \dot\rho=-i[H(\Omega),\rho]+\gamma\,\mathcal{D}[\sigma_-]\rho ,
  \label{eq:num-model}
\end{equation}
with $\mathcal{D}[L]\rho = L\rho L^\dagger-\tfrac12\{L^\dagger L,\rho\}$. In
units $\gamma=1$, at the critical drive $\Omega_0=\gamma/4$, the Liouvillian
$\mathcal{L}_0$ has spectrum $\{0,\,-\gamma/2,\,-3\gamma/4\}$. The
degenerate eigenvalue $-3\gamma/4$ has geometric multiplicity one, so
$\mathcal{L}_0$ carries a size-two Jordan block. This is an order-two
Liouvillian exceptional point.

We perturb the drive amplitude, \(\Omega=\Omega_0+\varepsilon\), so that
\(\mathcal L_\varepsilon=\mathcal L_0+\varepsilon\mathcal L_1\) with
\(\mathcal L_1=-i[\sigma_x/2,\,\cdot\,]\), and take
\(\rho(0)=|e\rangle\langle e|\). We use the finest partition of the nonzero
spectrum, \(S=\{x,J\}\): the \(\sigma_x\) mode, with eigenvalue \(-\gamma/2\)
and projector \(P_x\), and the Jordan sector, with eigenvalue \(-3\gamma/4\)
and projector \(P_J\). The stationary sector is labeled \(0\). The projectors
are obtained algebraically from the generalized eigenvectors of
\(\mathcal L_0\). All operator norms are taken as the Hilbert--Schmidt-induced
norm. The constants of Sec.~\ref{sec:error_bounds} are then
\begin{equation}
\begin{aligned}
  M_x &= 1.00, \qquad & M_J &= 1.41, \qquad & C_\rho &= 2.93,\\
  H_x &= 2.00, \qquad & H_J &= 3.10,
\end{aligned}
\label{eq:weights}
\end{equation}
where \(H_J\) is the closed form of Eq.~\eqref{eq:Ha}. The drive does not
couple the \(\sigma_x\) mode to the other sectors, so, with rows and columns
ordered as \((x,J,0)\),
\begin{equation}
  A=\begin{pmatrix} 0 & 0 & 0\\ 0 & 4.37 & 3.08\\ 0 & 0 & 0 \end{pmatrix},
\end{equation}
and \(R=\lVert A\rVert_1=r(A)=A_{JJ}=4.37\). With the exact integral of
Eq.~\eqref{eq:Ha} instead of the closed form, \(H_J=2.35\) and \(R=3.31\).
The coefficients \(\rho_k(t)\) are computed exactly as blocks of the
exponential of a block-bidiagonal matrix with \(\mathcal L_0\) on the diagonal
and \(\mathcal L_1\) on the superdiagonal~\cite{VanLoan1978}.

\begin{figure*}[t]
  \centering
  \includegraphics[width=\textwidth]{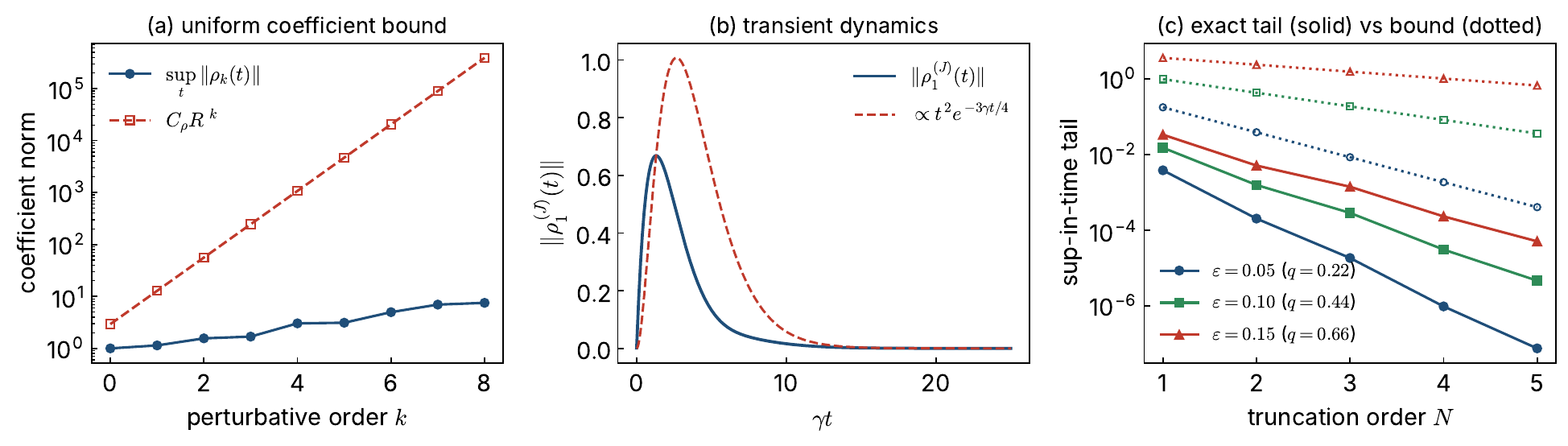}
\caption{\label{fig:ep-benchmark}%
  Validation at the single-ion Liouvillian exceptional point
  [Eq.~\eqref{eq:num-model}, $\Omega_0=\gamma/4$, $\gamma=1$].
  (a)~$\sup_t\lVert\rho_k(t)\rVert$ (blue) stays below the bound $C_\rho R^k$
  of Theorem~\ref{thm:uniform} (red) at every order.
  (b)~The defective single-block first-order contribution $\rho_1^{(J)}(t)$
  [Eq.~\eqref{eq:num-transient}] grows polynomially and decays as
  $e^{-3\gamma t/4}$. The highest-degree $t^{2m-1}=t^{3}$ term of
  Proposition~\ref{prop:jordan} is present but numerically subdominant, so the
  curve is nearly linear on the rising side, and it vanishes at long times
  (Sec.~\ref{sec:transient_amplification}).
  (c)~The exact sup-in-time tail after order $N$ (solid) stays below the
  geometric bound $C_\rho q^{N+1}/(1-q)$ (dotted) for three perturbation
  strengths.}
\end{figure*}

\subsection{\label{sec:numerics-bounds}Uniform bounds and transient response}

The three panels of Fig.~\ref{fig:ep-benchmark} confirm the estimates.
Panel~(a): the exact coefficient norm obeys
$\sup_t\lVert\rho_k(t)\rVert\le C_\rho R^k$ at every order. Panel~(c): the
tail bound of Theorem~\ref{thm:uniform}\rev{, Eq.~\eqref{eq:tail_q},} holds
for $\varepsilon=0.05,0.10,0.15$ ($q=0.22,0.44,0.66$), and the actual tail
falls geometrically below it.

Panel~(b) tests the single-block scaling. The first-order contribution of the
defective block is a degree-$(2m-1)$ polynomial times the block exponential,
\begin{equation}
  \begin{split}
    \rho_1^{(J)}(t)=e^{\lambda t}\Big[
      &\,t\,P_J\mathcal{L}_1P_J\,\rho(0)\\
      &+\tfrac{t^2}{2}\bigl(P_J\mathcal{L}_1N_m+N_m\mathcal{L}_1P_J\bigr)\rho(0)\\
      &+\tfrac{t^3}{6}\,N_m\mathcal{L}_1N_m\,\rho(0)\Big],
  \end{split}
  \label{eq:num-transient}
\end{equation}
with $\lambda=-3\gamma/4$, block size $m=2$, and $P_J$, $N_m$ the projection and
nilpotent of the defective block (the $N_m$ of
Sec.~\ref{sec:exceptional_points}, satisfying $N_m^2=0$ here). The three
coefficients of Eq.~\eqref{eq:num-transient} have norms $1.37$, $0.34$, and
$0.05$, so the highest-degree $t^{2m-1}=t^3$ term of
Proposition~\ref{prop:jordan} is present, though numerically subdominant: it
is roughly $28$ times smaller than the linear term, and $\rho_1^{(J)}(t)$
appears nearly linear on its rising side. \rev{This is consistent with
$\eta=\lVert N_mP_J\rVert/(\lVert P_J\rVert\gamma_J)=0.65<1$
(Sec.~\ref{sec:transient_amplification}), with $\gamma_J=3\gamma/4$: during the
lifetime of the mode, the Jordan terms stay smaller than the exponential term.
Here $\lVert N_mP_J\rVert=0.687$ is the spectral response strength of the
exceptional point.} The response vanishes as $t\to\infty$, since the block is
nonperipheral (\rev{Sec.~\ref{sec:transient_amplification}}).

\rev{The transient part of $\rho_k$ also grows faster at high order than its
stationary limit, Eq.~\eqref{eq:drazin}: for $k\in[50,60]$, the growth factors
per order are $1.38$ for the transient part and $1.33$ for the stationary
limit. Finite-time quantities are therefore not fixed by the stationary series
of Ref.~\cite{NagibWalker2025}.}

\begin{revblock}
\subsection{Symmetry-breaking perturbations}
\label{sec:numerics-symmetry}
In the Bloch representation, \(\rho=(I+x\sigma_x+y\sigma_y+z\sigma_z)/2\), the
model of Eq.~\eqref{eq:num-model} reads
\(\dot x=-\gamma x/2\), \(\dot y=-\gamma y/2-\Omega z\), and
\(\dot z=\Omega y-\gamma(z+1)\). The \(x\) component is decoupled: the equations
are invariant under \(x\to-x\), and the \(\sigma_x\) mode, which decays at rate
\(\gamma/2\), never exchanges weight with the \((y,z)\) sector that contains the
Jordan block and the steady state. A change of the drive amplitude preserves
this symmetry, so in Secs.~\ref{sec:numerics-model}
and~\ref{sec:numerics-bounds} the perturbation never couples the \(\sigma_x\)
mode to the rest of the dynamics. To test the general situation, we use two
perturbations that break the symmetry and couple the block to the
\(\sigma_x\) mode: a detuning, \(\mathcal L_1=-i[\sigma_z/2,\,\cdot\,]\), which
adds \(\dot x=-y\) and \(\dot y=x\), and a dephasing along the axis
\((1,0,1)/\sqrt2\), \(\mathcal L_1=\mathcal D[(\sigma_x+\sigma_z)/\sqrt2]\),
which couples \(x\) and \(z\) and damps \(y\). The first is Hamiltonian and the
second dissipative. The drive and the tilted dephasing split the exceptional pair as
\(\varepsilon^{1/2}\), the generic behavior at an exceptional point of order
two~\cite{Seyranian_2005}, whereas the detuning, for which
\(P_J\mathcal L_1P_J=0\), splits it linearly.

Table~\ref{tab:radii} collects the constants for the three perturbations. For
the detuning, the perturbation connects only the $x$ and $J$ sectors, so
$r(A)=\sqrt{A_{xJ}A_{Jx}}=2.61$ lies below $R=3.10$. The refined tail bound
of Eq.~\eqref{eq:tail_refined} holds in every case tested with
$|\varepsilon|\,r(A)<1$, including one with $q>1$.

\begin{table}[t]
  \revfloat
  \caption{\label{tab:radii}%
    Growth constants per order at the exceptional point. $R$ and $r(A)$ use
    the closed-form weights of Eq.~\eqref{eq:Ha}, $r(A)^{\rm ex}$ the exact
    integrals. The last column is the observed rate
    $(\sup_t\lVert\rho_{30}\rVert)^{1/30}$. $C_\rho=2.93$ in all cases.}
  \begin{tabular}{lcccc}
    \toprule
     & $R$ & $r(A)$ & $r(A)^{\rm ex}$ & rate$_{30}$\\
    \midrule
    drive & 4.37 & 4.37 & 3.31 & 1.35\\
    detuning & 3.10 & 2.61 & 2.27 & 1.82\\
    tilted & 9.60 & 8.25 & 6.60 & 3.26\\
    \bottomrule
  \end{tabular}
\end{table}

We also tested the frequency-domain bound of Eq.~\eqref{eq:freq_bound} and the
time-dependent bound of Sec.~\ref{sec:time_dependent} for the three
perturbations, with the exact weights $H_a$ and orders $k\le6$. In the
frequency domain, we took $s=i\omega$ at $800$ frequencies with
$10^{-3}\le|\omega|/\gamma\le10^{2}$, on the boundary of the half-plane where
Eq.~\eqref{eq:freq_bound} holds. For the time-dependent case, we took
$\mathcal L_1(t)=\cos(\nu t)\,\mathcal L_1$ with
$\nu/\gamma\in\{0.2,0.5,1,2\}$, for which the constants are those of the
static perturbation. Both bounds held in every case, with largest ratios
between the exact quantity and the bound of $0.70$ in the frequency domain
and $0.12$ in the time domain.

\end{revblock}

\subsection{\label{sec:numerics-sharpness}Sharpness of the effective parameter}

\rev{At an exceptional point, the bound of Theorem~\ref{thm:uniform} is
conservative, because the generator is strongly non-normal and norm estimates
cannot capture cancellations between paths.} To test the opposite regime, in
which the bound is sharp, we take a diagonalizable model with no exceptional
point: amplitude damping perturbed by a pure detuning,
\begin{equation}
  \mathcal{L}_0=\gamma\,\mathcal{D}[\sigma_-],\qquad
  \mathcal{L}_1=-i[\sigma_z/2,\,\cdot\,],
  \label{eq:num-tight}
\end{equation}
with $\rho(0)=\lvert +\rangle\!\langle +\rvert$ and, unlike in
Sec.~\ref{sec:numerics-model}, $\gamma=1/2$. The perturbation acts diagonally
on the coherence eigenmodes $|e\rangle\langle g|$ and $|g\rangle\langle e|$,
which decay at rate $\gamma/2$, so the block paths reduce to a single
self-loop and the coefficient has the closed form
\begin{equation}
  \sup_{t\ge0}\lVert\rho_k(t)\rVert
   = c_0\,R^{k}\,\frac{k^{k}e^{-k}}{k!}
   \;\xrightarrow[k\to\infty]{}\; \frac{c_0\,R^{k}}{\sqrt{2\pi k}},
   \label{eq:num-saturation}
\end{equation}
with $R=2/\gamma=4$ and $c_0=\lVert P_{\mathrm{coh}}\rho(0)\rVert=1/\sqrt2$,
where $P_{\mathrm{coh}}$ is the spectral projection of $\mathcal L_0$ onto the
coherence modes. The path bound $\mathbf 1^{T}A^ky=c_0R^k$ of
Eq.~\eqref{eq:path_bound} is thus saturated up to a factor
$(2\pi k)^{-1/2}$, and $R$ is the exact growth rate,
$\limsup_k(\sup_t\lVert\rho_k\rVert)^{1/k}=R$.

\begin{figure*}[t]
  \revfloat
  \centering
  \includegraphics[width=\textwidth]{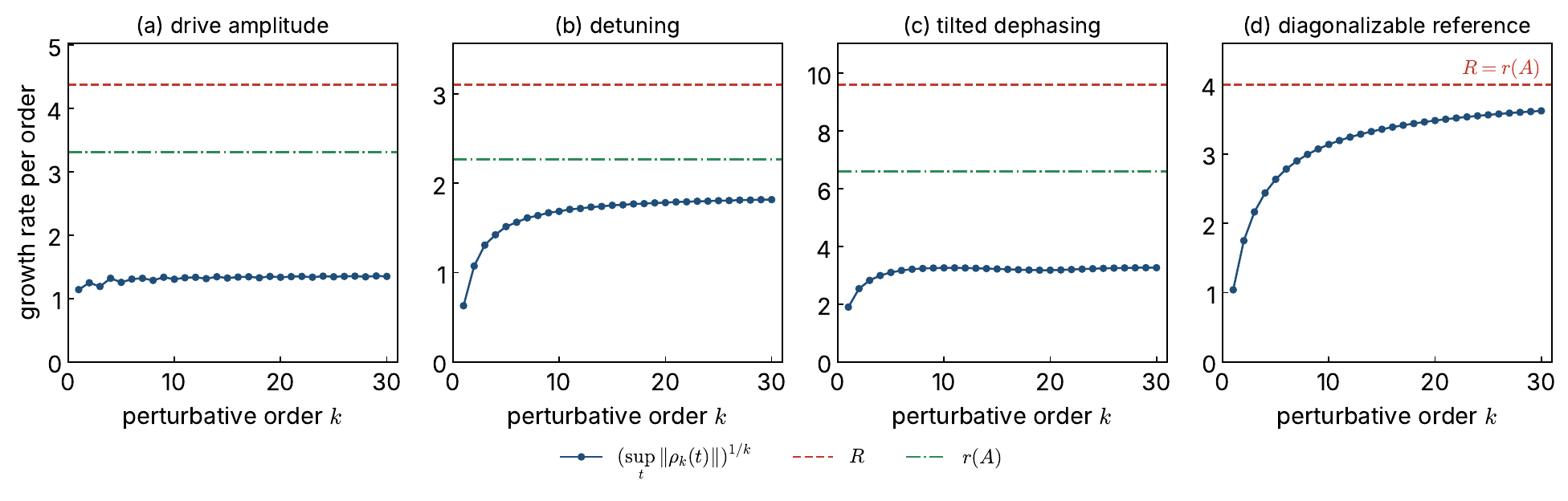}
\caption{\label{fig:rates}%
  Growth rate per order, $(\sup_t\lVert\rho_k(t)\rVert)^{1/k}$, for $k\leq30$
  (markers), compared with $R$ of Theorem~\ref{thm:uniform}, computed with the
  closed-form weights of Eq.~\eqref{eq:Ha} (dashed), and with $r(A)$ computed
  with the exact weights $H_a$ (dash-dotted). (a)--(c)~Exceptional-point model
  [Eq.~\eqref{eq:num-model}, $\Omega_0=\gamma/4$, $\gamma=1$] under the drive
  and the two symmetry-breaking perturbations of
  Sec.~\ref{sec:numerics-symmetry}. (d)~Diagonalizable reference model
  [Eq.~\eqref{eq:num-tight}, $\gamma=1/2$], where $r(A)=R=4$.}
\end{figure*}

\rev{Figure~\ref{fig:rates} compares the growth rate per order,
\((\sup_t\lVert\rho_k\rVert)^{1/k}\), in the two models. In the reference model
it approaches \(R=4\) [Fig.~\ref{fig:rates}(d)]. In the exceptional-point
model, the exact coefficients obey Theorem~\ref{thm:uniform} at all orders
computed (\(k\leq30\)) for the three perturbations, and the rate stays well
below \(R\) [Figs.~\ref{fig:rates}(a)--\ref{fig:rates}(c)]. Under the drive it
saturates near \(1.35\), about a third of \(R=4.37\), which measures how
conservative the bound is at the exceptional point.}

\begin{revblock}
\subsection{Sweep through the exceptional point}
\label{sec:numerics-sweep}

Theorem~\ref{thm:EP} concerns a family $\mathcal L_0(\delta)$ rather than a
fixed generator. Figure~\ref{fig:sweep} tests it along
$\Omega=\Omega_0+\delta$ (in units $\gamma=1$) with two partitions of the
spectrum. Panel~(a) shows the constants that do not depend on the
perturbation. With the separate eigenprojectors $P_\pm(\delta)$, these
constants diverge as the exceptional point is approached: $H_\pm$ and
$M_\pm=\lVert P_\pm\rVert$ grow as $|\delta|^{-1/2}$, and $C_\rho$ as
$|\delta|^{-1}$. The prefactor agrees with Eq.~\eqref{eq:Ppm}: with
$|\mu|\simeq\sqrt{|\delta|/2}$,
$\lVert P_+\rVert\sqrt{2|\delta|}\to\lVert N_mP_J\rVert=0.687$. In
panels~(b)--(d), $R$ diverges as $|\delta|^{-3/2}$ for the drive and the
tilted dephasing, where three factors of order $|\delta|^{-1/2}$ enter
$H_\pm\lVert P_\pm\mathcal L_1P_\pm\rVert$, and as $|\delta|^{-1}$ for the
detuning, where $P_c\mathcal L_1P_c=0$. With the cluster projector
$P_c=P_++P_-$, all constants are continuous through the exceptional point and
reach the Jordan-sector values at $\delta=0$. Panels~(e)--(h) show that they
vary smoothly, by less than $6\%$ over $|\delta|\leq0.1$.

\vfill

\begin{figure*}[t]
  \revfloat
  \centering
  \includegraphics[width=\textwidth]{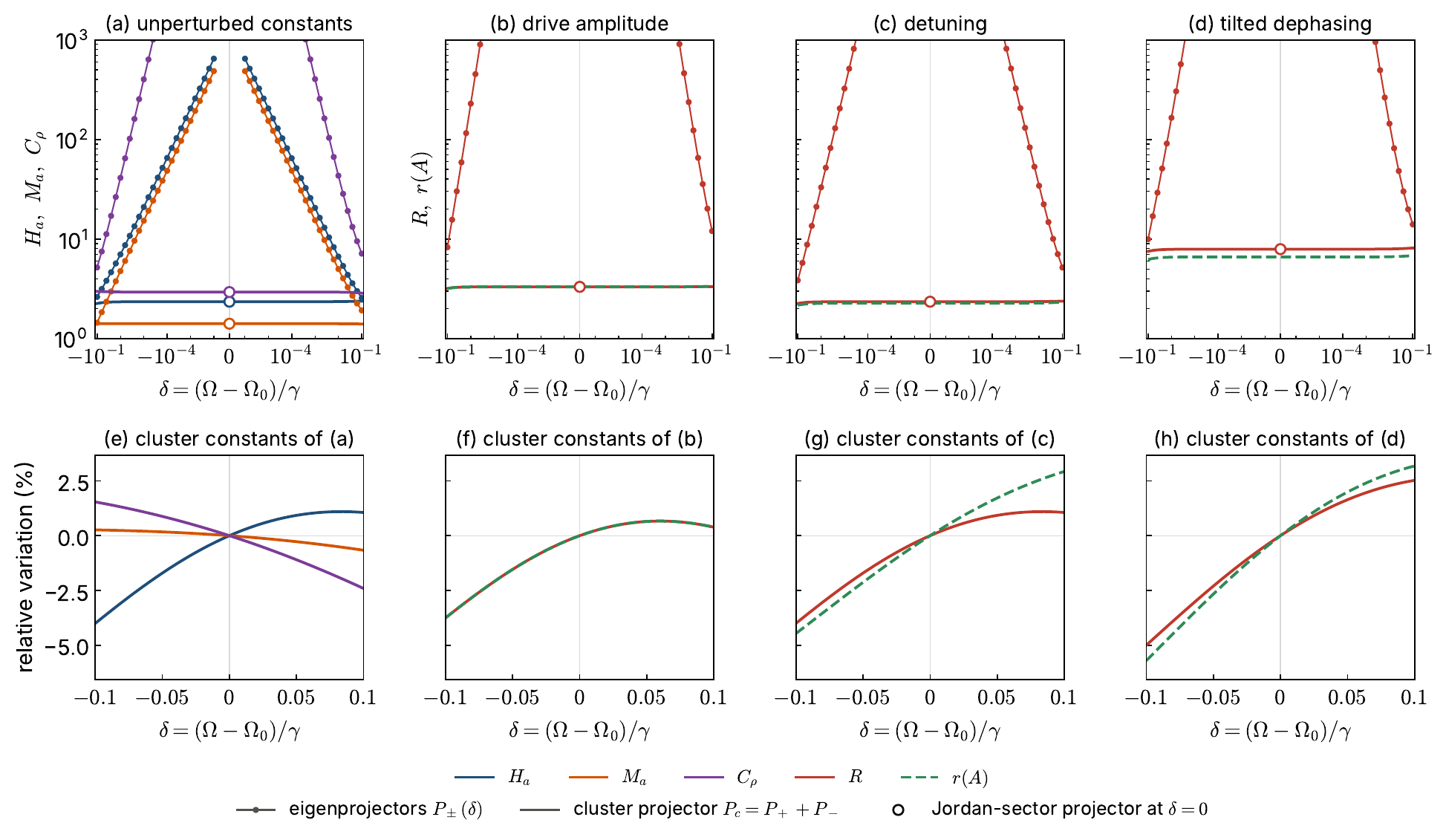}
\caption{\label{fig:sweep}%
  Constants of Theorem~\ref{thm:uniform} as functions of the distance from
  the exceptional point, $\delta=(\Omega-\Omega_0)/\gamma$ (symmetric
  logarithmic axis, linear for $|\delta|<10^{-6}$), computed with the exact
  weights $H_a$. (a)~Constants that do not depend on the perturbation.
  (b)--(d)~$R$ and $r(A)$ for the three perturbations. With separate
  eigenprojectors $P_\pm(\delta)$ (dots), $H_\pm$ and $M_\pm=\lVert P_\pm\rVert$
  diverge as $|\delta|^{-1/2}$, $C_\rho$ as $|\delta|^{-1}$, and $R$ as
  $|\delta|^{-3/2}$, or $|\delta|^{-1}$ for the detuning. With the cluster
  projector $P_c=P_++P_-$ (solid), all constants are continuous and reach the
  Jordan-sector values (open circles) at $\delta=0$ (Theorem~\ref{thm:EP}).
  For the drive, $R$ and $r(A)$ coincide. (e)--(h)~Relative variation
  $[X(\delta)-X(0)]/X(0)$ of the cluster constants of the panels above, on a
  linear axis.}
\end{figure*}

\end{revblock}

\section{\label{sec:conclusion}Conclusion}

In this paper, we used a recursive time-domain perturbative approach to study
the response of finite-dimensional open quantum systems \rev{at, }near\rev{,
and away from} Liouvillian exceptional
points~\cite{Heiss_2012,Miri_2019,Ashida_2020,Zhou2023AcceleratingRelaxation}.
\rev{We used the Dyson--Duhamel expansion to calculate the perturbative
corrections to the density matrix.} A central feature of the formulation is
its behavior near non-diagonalizable generators. When the unperturbed
Lindbladian contains a Jordan block, the associated polynomial-in-time factors
are already present in the propagator and are therefore inherited by the
perturbative corrections. For nonperipheral Jordan blocks, these polynomial
factors multiply exponentially decaying terms. Thus, defective Liouvillian
modes can enhance transient responses while remaining asymptotically stable,
since the corresponding eigenvalues have negative real
part~\cite{Davies_1974,Breuer_2002}. \rev{Polynomial factors also arise for
diagonalizable generators, as resonances along perturbative paths. We
quantified the excess degree produced by Jordan blocks and bounded the
perturbative contributions confined to Jordan sectors with explicit
constants.}

We also derived time-uniform estimates for the perturbative coefficients
under three assumptions: a unique stationary state, no other peripheral
eigenvalues, and trace-preserving
perturbations~\cite{Gorini_1976,Lindblad_1976,Breuer_2002}. Under these
assumptions, secular contributions associated with repeated stationary
projections are removed by the identity $P_0\mathcal L_1=0$. The resulting
expansion is controlled by an effective parameter \(q=|\varepsilon|R\), where
\(R\) \rev{is computed from unperturbed data alone and} depends on the
spectral gaps, the Jordan-block structure, and the block-resolved matrix
elements of the perturbation. Hence, the estimates make explicit how small
\(\varepsilon\) must be for the perturbative coefficients to be uniformly
controlled in time. \rev{The spectral radius of the full comparison matrix
sharpens the estimate when the perturbation connects only some spectral
sectors. Keeping the eigenvalues that coalesce at an exceptional point in one
cluster makes all constants continuous through it, so the estimates are also
uniform in the distance from the exceptional point. The same bound covers
time-dependent perturbations of a time-independent generator and the
frequency-domain response.}

\rev{We tested the estimates in a driven two-level system with a Liouvillian
exceptional point, under perturbations that preserve or break its symmetry,
and along a sweep through the exceptional point. The bounds held in every
case. At the exceptional point, they overestimate the growth rate per order
by up to a factor of about three, whereas in a diagonalizable reference model
\(R\) is the exact growth rate.}

Possible applications for our estimatives include local parameter sensitivity, driven-dissipative
control and perturbative corrections to quantum thermodynamic
cycles~\cite{Alicki_1979,Kosloff_2013,Vinjanampathy_2016,Goold2016,Binder_2018,Myers2022}. \rev{Natural directions for future work are sharper a
priori constants, which would account for cancellations between perturbative
paths, and extensions to time-dependent unperturbed generators and to
generators with several stationary states or oscillating peripheral modes.}

\begin{acknowledgments}
This study was financed in part by the Coordenação de Aperfeiçoamento de Pessoal de Nível Superior – Brazil (CAPES) – Finance Code 001. Claude Opus 5.5 (Anthropic) was used to help write the code for the simulations and figures in this paper, and to improve the clarity of the exposition.
\end{acknowledgments}

\bibliographystyle{apsrev4-2}
\bibliography{references}

\end{document}